\documentclass[prl, twocolumn, showpacs]{revtex4}
\usepackage{amsfonts}
\usepackage{graphicx}
\usepackage{epsfig}
\begin{document}
\title{Gluino Stransverse Mass
}
\author{Won Sang Cho$^1$, Kiwoon Choi$^1$, Yeong Gyun Kim$^{1,2}$ and Chan Beom Park$^1$ }
\affiliation{ $^1$ {\it Department of Physics, KAIST, Daejon 305--017, Korea} \\
$^2$ {\it ARCSEC, Sejong University, Seoul 143-747, Korea}
}

\begin{abstract}
We introduce a new observable, 'gluino stransverse mass', which is
an application of the Cambridge $m_{T2}$ variable to the process
where gluinos are pair produced in proton-proton collision and
each gluino subsequently decays into two quarks and one LSP,
$i.e.~ \tilde{g}\tilde{g} \rightarrow qq\tilde\chi_1^0\
qq\tilde\chi_1^0$. We show that the gluino stransverse mass can be
utilized to measure the gluino mass and the lightest neutralino
mass separately, and also the 1st and 2nd generation squark masses
if squarks are lighter than gluino, thereby providing a good first
look at  the pattern of sparticle masses experimentally.
\end{abstract}
 \maketitle
The Large Hadron Collider (LHC) at CERN  will soon explore TeV
energy scale, where new physics beyond the Standard Model (SM)
likely reveals itself \cite{atlas,cms}. Among various new physics
proposals, weak scale supersymmetry (SUSY) \cite{susy} is perhaps
the most promising candidate, which provides a solution for gauge
hierarchy problem and complies with gauge coupling unification.
Furthermore, with R-parity conservation, the lightest
supersymmetric particle (LSP) becomes a natural candidate for the
non-baryonic dark matter in the Universe.

Once SUSY signals are discovered through event excess beyond the SM
backgrounds in inclusive search channels, the next step will be the
measurements of SUSY particle masses and their physical properties
in various exclusive decay chains. Then it might be possible to
reconstruct SUSY theory, in particular SUSY breaking mechanism from
the experimental information on SUSY particle masses. In this
regard, determination of gaugino masses has a significant
implication because predictions of the gaugino masses are rather
robust compared to those on sfermion masses  \cite{choi-nilles}.

In this paper, we introduce a new observable, `gluino stransverse
mass', which is an application of $m_{T2}$ variable \cite{lester} to
the process where gluinos are pair produced in proton-proton
collision and each gluino subsequently decays into two quarks and
one LSP, \begin{eqnarray}
 pp\rightarrow \tilde{g}\tilde{g} \rightarrow
qq\tilde\chi_1^0\ qq\tilde\chi_1^0,\end{eqnarray} where $q$ stands
for the 1st or 2nd generation quark.
 We show that the gluino
stransverse mass can be utilized to measure the gluino mass and the
lightest neutralino mass separately, and also the 1st and 2nd
generation squark masses if squarks are lighter than gluino, thereby
providing a good first look at the pattern of sparticle masses
experimentally.



If light enough, gluinos would be  pair produced copiously in
proton-proton collision ($pp\rightarrow\tilde g\tilde g$),
 and
each gluino decays into two quarks and one LSP ($\tilde g
\rightarrow q q \tilde\chi_1^0$) through three-body decay induced
by an exchange of off-shell squark or two body cascade decay with
intermediate on-shell squark.
For each gluino decay $\tilde g\rightarrow  qq\tilde\chi_1^0$, a
transverse mass is constructed, which is defined as
\begin{eqnarray}
m_{T}^2 (m_{qqT}, m_\chi, \bold{p}_T^{qq}, \bold{p}_T^{\chi})= m_{qqT}^2 + m_\chi^2 \nonumber \\
+ 2 (E_T^{qq} E_T^{\chi}-\bold{p}_T^{qq}\cdot\bold{p}_T^{\chi}),
\end{eqnarray}
where $m_{qqT}$ and $\bold{p}_T^{qq}$ are the transverse invariant
mass and transverse momentum of the $qq$ system, respectively,
while $m_\chi$ and $\bold{p}_T^{\chi}$ are the {\it assumed} mass
and transverse momentum of the LSP, respectively. The transverse
energies of the $qq$ system and of the LSP are defined by
\begin{eqnarray}
E_T^{qq} \equiv \sqrt{|{\bold p_T^{qq}}|^2 + m_{qqT}^2},\quad
E_T^{\chi} \equiv \sqrt{|{\bold p_T^{\chi}}|^2 + m_\chi^2}.
\end{eqnarray}
 With two  such gluino decays in each event, the gluino
stransverse mass $m_{T2} (\tilde g)$ is defined as
\begin{eqnarray}
m_{T2}^2 (\tilde g) \equiv \min_{{\bf p}_{T}^{\chi(1)}+{\bf p}_{T}^{\chi(2)}={\bf p}_T^{miss}}
\left[ {\rm max} \{ m_T^{2(1)}, m_T^{2(2)} \} \right],
\label{eq:mt2def}
\end{eqnarray}
where the minimization is performed over all possible splitting of
the observed missing transverse energy $\bold{p}_T^{miss}$ into
two assumed transverse momenta, $\bold{p}_T^{\chi(1)}$ and
$\bold{p}_T^{\chi(2)}$.

From the definition (\ref{eq:mt2def}) of the gluino stransverse
mass, one obtains a relation:
\begin{eqnarray}
m_{T2} (\tilde g) \leq m_{\tilde g} ~~~{\rm for}~~~
m_\chi=m_{\tilde\chi_1^0},
\end{eqnarray}
i.e. $m_{T2} (\tilde g)$ is less than or equal to the true gluino
mass $m_{\tilde g}$ if the {\it trial} LSP mass $m_\chi$ is equal to
the {\it true} LSP mass $m_{\tilde\chi_1^0}$. Therefore one can
determine $m_{\tilde g}$ from the endpoint measurement of $m_{T2}
(\tilde g)$ distribution: \begin{eqnarray} m_{T2}^{\rm max} (m_\chi)
\equiv \max_{\mbox{all events}}\left[
m_{T2}(\tilde{g})\right],\end{eqnarray}
 if we already know the true LSP mass.
However $m_{\tilde\chi_1^0}$  might not be known in advance and
then, for a trial $m_\chi$  which is different from
$m_{\tilde\chi_1^0}$, $m_{T2}^{\rm max} (m_\chi)$ will differ from
$m_{\tilde g}$ and can be considered as  a function of $m_\chi$.

As we will see,  $m_{T2}^{\rm max}(m_\chi)$ has different
functional form depending upon whether the 1st and 2nd generation
squarks are heavier or lighter than gluino.
Thus, in order to investigate the $m_\chi$-dependence of $m^{\rm
max}_{T2}$,
we consider two cases separately.

If squark masses are larger than gluino mass, the gluino will
undergo three body decay into two quarks and an invisible LSP
through the exchange of an off-shell squark:
\begin{eqnarray}
\tilde g \rightarrow q q \tilde\chi_1^0.
\end{eqnarray}
In order to get an idea on how $m_{T2}^{\rm max} (m_\chi)$ are
determined for a generic value of the trial LSP mass $m_\chi$, we
consider two extreme momentum configurations and then construct the
 gluino stransverse mass associated with each of them.

The first momentum configuration  is that two gluinos are produced
at rest and then each gluino subsequently decays into two quarks
moving in the same direction, and one LSP whose direction is
opposite to the quark direction. Furthermore, two sets of gluino
decay products are parallel to each other and all of them are on the
transverse plane with respect to the proton beam direction. For such
momentum configuration, we have
\begin{eqnarray}
m_{qqT}^{(1)}=m_{qqT}^{(2)}=0,\end{eqnarray} where the final state
quarks are regarded to be {\it massless}. The transverse energies
and transverse momenta of the $qq$ systems are given by
\begin{eqnarray}
\label{etqq} &&  E_T^{qq(1)}=E_T^{qq(2)}
= |\bold{p}_T^{qq(1)}| =|\bold{p}_T^{qq(2)}|\nonumber \\
&&= {m_{\tilde g}^2 -m_{\tilde\chi_1^0}^2 \over 2 m_{\tilde
g}}\,\equiv\, E_T^{qq}(max),
\end{eqnarray}
where $m_{\tilde g}$ and $m_{\tilde\chi_1^0}$ are the true gluino
mass  and the true LSP mass, respectively, and the corresponding
total missing transverse momentum is
\begin{eqnarray}
|\bold{p}_T^{miss}|=2 E_T^{qq}(max). \label{ptmax}
\end{eqnarray}

It has been shown that for certain momentum configurations, the
$m_{T2}$ variable (`balanced solution') can be obtained as the
minimum of $m_T^{(1)}$ subject to the following two constraints
\cite{lester}:
\begin{eqnarray}
m_T^{(1)}=m_T^{(2)},\quad
\bold{p}_T^{miss}=\bold{p}_T^{\chi(1)}+\bold{p}_T^{\chi(2)}.
\end{eqnarray}
If one applies this balanced solution approach to the momentum
configurations with the above $m_{qqT}^{(1,2)}$, $E_T^{qq(1,2)}$, 
$\bold{p}_T^{qq(1,2)}$, $\bold{p}_T^{miss}$
and still undetermined $\bold{p}_T^{\chi(1,2)}$,
one finds that the minimum of $m_T^{(1)}(=m_T^{(2)})$ is obtained when
$\bold{p}_T^{\chi(1)}=\bold{p}_T^{\chi(2)}=\bold{p}_T^{miss}/2$,
leading to the following  gluino stransverse mass:
\begin{eqnarray}
m_{T2} (\tilde g)= E_T^{qq} (max) + \sqrt{(E_T^{qq} (max))^2 +
m_\chi^2} \label{eq:mg1}
\end{eqnarray}
for  generic $m_\chi$.
One can show that this $m_{T2}(\tilde{g})$  corresponds to
$m_{T2}^{\rm max}$ for $m_\chi\leq m_{\tilde\chi_1^0}$ \cite{cho}:
\begin{eqnarray}
m^{\rm max}_{T2}(m_\chi)&=& {m_{\tilde g}^2 -m_{\tilde\chi_1^0}^2
\over 2 m_{\tilde g}} + \sqrt{\left({m_{\tilde g}^2
-m_{\tilde\chi_1^0}^2
\over 2 m_{\tilde g}}\right)^2 + m_\chi^2} \nonumber \\
&&\mbox{for}\quad m_\chi\leq m_{\tilde\chi_1^0} .
\label{curve1}\end{eqnarray} Note that $m^{\rm max}_{T2}
(m_\chi=m_{\tilde\chi_1^0})=m_{\tilde g}$ as required.

Other extreme momentum configuration which would determine
$m_{T2}^{\rm max}(m_\chi)$ for $m_\chi\geq m_{\tilde\chi_1^0}$  is
that gluinos are pair produced at rest and for each gluino decay,
two quarks are back to back to each other while LSP is at rest. In
addition,
all particles are on the transverse plane.
In this case, one easily finds
\begin{eqnarray}
 m_{qqT}^{(1)}=m_{qqT}^{(2)}=m_{\tilde
g}-m_{\tilde\chi_1^0}\,\equiv\, m_{qqT}(max), \label{mqqt}
\end{eqnarray}
and also 
\begin{eqnarray} 
E_T^{qq(1)}=E_T^{qq(2)}=m_{qqT}(max),
\label{eqqt}
\end{eqnarray} with
\begin{eqnarray} 
{\bold p_T^{qq(1)}}={\bold p_T^{qq(2)}}={\bold p_T^{miss}}=0. 
\label{pqqt}
\end{eqnarray} 

For the momentum configurations with $m_{qqT}^{(1,2)}$, $E_T^{qq(1,2)}$,
$\bold{p}_T^{qq(1,2)}$ and $\bold{p}_T^{miss}$ given by
(\ref{mqqt}), (\ref{eqqt}), (\ref{pqqt}) respectively, 
$m_T^{(1)}$ is  equal to $m_T^{(2)}$ for all possible
splitting of the missing transverse momentum:
$\bold{p}_T^{miss}=0=\bold{p}_T^{\chi(1)}+\bold{p}_T^{\chi(2)}$, and
the minimum of $m_T^{(1)} (=m_T^{(2)})$ occurs when
$\bold{p}_T^{\chi(1)}=\bold{p}_T^{\chi(2)}=0$.
Then the gluino stransverse mass obtained as a balanced solution is
given by
\begin{eqnarray}
m_{T2} ({\tilde g}) = m_{qqT} (max) + m_\chi
\label{eq:mg2}
\end{eqnarray}
for generic value of $m_\chi$. This in fact corresponds to
$m_{T2}^{\rm max}$ for $m_\chi\geq m_{\tilde\chi_1^0}$ \cite{cho}:
\begin{eqnarray}
m_{T2}^{\rm
max}(m_\chi)&=&\left(m_{\tilde{g}}-m_{\tilde\chi_1^0}\right)+m_\chi
\nonumber \\
&&\mbox{for}\quad m_\chi\geq m_{\tilde\chi_1^0},
\label{curve2}\end{eqnarray} which again gives $m^{\rm max}_{T2}
(m_\chi=m_{\tilde\chi_1^0})=m_{\tilde g}$.

The above momentum configurations leading to the expression
(\ref{curve2}) have $\bold{p}_T^{miss}=0$, thus could be eliminated
by the event cut imposing a lower bound on $|\bold{p}_T^{miss}|$ in
the real data analysis. However, a more detailed study \cite{cho}
shows that there exist momentum configurations yielding the same
expression of $m^{\rm max}_{T2}(m_\chi)$ for $m_\chi\geq
m_{\tilde{\chi}_1}^0$, while having a {\it sizable}
$|\bold{p}_T^{miss}|$ comparable to $m_{\tilde{g}}/2$, for instance
a configuration in which the two quarks from the first gluino move
in the same transverse direction, while the other two quarks from
the second gluino are back to back. As a result, the functional
behavior of (\ref{curve2}) can still be constructed from collider
data under a proper cut on the missing transverse momentum.

By now, it should be clear that $m^{\rm max}_{T2}$  for
$m_\chi<m_{\tilde\chi_1^0}$  (Eq. (\ref{curve1})) has a quite
different form from $m^{\rm max}_{T2}$  for $m_\chi >
m_{\tilde\chi_1^0}$ (Eq. (\ref{curve2})). As required, they should
cross at $m_\chi = m_{\tilde\chi_1^0}$.
{\it Thus, if the function $m^{\rm max}_{T2}(m_\chi)$ could be
constructed from experimental data, which would identify  the
crossing point, one will be able to determine the gluino mass and
the LSP mass simultaneously.}



The experimental feasibility of measuring $m_{\tilde{g}}$ and
$m_{\tilde{\chi}^0_1}$ through $m^{\rm max}_{T2}$ depends on the
systematic uncertainty associated with the jet resolution since
$m^{\rm max}_{T2}$ is obtained mostly from the momentum
configurations in which some (or all) quarks move in the same
direction. Our Monte Carlo study  indicates that the resulting error
is not so significant, so that $m_{\tilde{g}}$ and
$m_{\tilde{\chi}^0_1}$ can be determined rather accurately by the
crossing behavior of $m^{\rm max}_{T2}$. As a specific example, we
have examined a parameter point in the minimal anomaly mediated
SUSY-breaking (mAMSB) scenario \cite{amsb} with heavy squarks, which
gives
\begin{eqnarray}
m_{\tilde g} = 780.3 \ {\rm GeV},\ m_{\tilde\chi_1^0} = 97.9 \
{\rm GeV}, \nonumber
\end{eqnarray}
and a few TeV masses for sfermions.
We have generated a Monte Carlo sample of SUSY events for
proton-proton collision at 14 TeV by PYTHIA \cite{pythia}. The event
sample corresponds to 300 $fb^{-1}$ integrated luminosity. We have
also generated SM backgrounds such as $t\bar{t}, W/Z+jet, WW/WZ/ZZ$
and QCD events, with less equivalent luminosity. The generated
events have been further processed with a modified version of fast
detector simulation program PGS \cite{pgs}, which approximates an
ATLAS or CMS-like detector with reasonable efficiencies and fake
rates.

The following event selection cuts are applied to have a clean
signal sample for gluino stransverse mass:
\begin{enumerate}
\item At least 4 jets with $P_{T1,2,3,4} > 200, 150, 100, 50$ GeV.
\item Missing transverse energy $E_T^{miss} > 250$ GeV. \item
Transverse sphericity $S_T > 0.25$. \item No b-jets and no
leptons.
\end{enumerate}
For each event, the four leading jets are used to calculate the
gluino stransverse mass. The four jets are divided into two groups
of dijets as follows. The highest momentum jet and the other jet
which has the largest $|p_{jet}| \Delta R$  with respect to the
leading jet are chosen as the two `seed' jets for division. Here
$p_{jet}$ is the jet momentum and $\Delta R \equiv
\sqrt{\Delta\phi^2+\Delta\eta^2}$, i.e. a separation in azimuthal
angle and pseudorapidity plane. Each of the remaining two jets is
associated to a seed jet which makes the smallest opening angle.
Then, each group of the dijets is considered to be originating
from the same mother particle (gluino).


%
%

\begin{figure}[ht!]
\begin{center}
\epsfig{figure=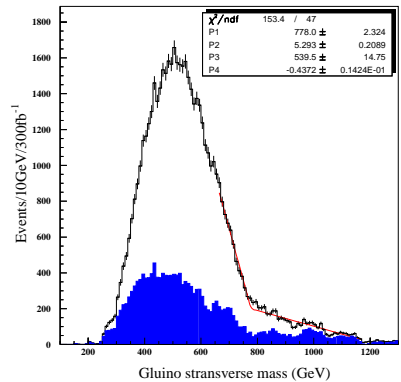,width=6cm,height=6cm}
\end{center}
\caption{\it The $m_{T2} (\tilde g)$ distribution with  $m_\chi=90$
GeV for the benchmark point of mAMSB with heavy squarks. Blue
histogram is the SM background.} \label{fig:amsb1}
\end{figure}

\begin{figure}[ht!]
\begin{center}
\epsfig{figure=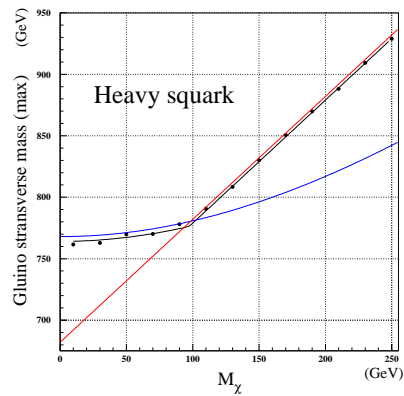,width=6cm,height=6cm}
\end{center}
\caption{\it $m_{T2}^{\rm max}$ as a function of the trial LSP mass
$m_\chi$ for the benchmark point of mAMSB with heavy squarks. }
\label{fig:amsb2}
\end{figure}

Fig.\ref{fig:amsb1} shows the resulting distribution of the gluino
stransverse mass for the trial LSP mass $m_\chi=90$ GeV. The blue
histogram corresponds to the SM background. Fitting with a linear
function with a linear background, we get the endpoint $778.0 \pm
2.3$ GeV. The measured edge values of $m_{T2} (\tilde g)$, i.e.
$m^{\rm max}_{T2}$, as a function of $m_\chi$ is shown in
Fig.\ref{fig:amsb2}.
Blue and red lines denote the theoretical curves of (\ref{curve1})
and (\ref{curve2}), respectively, which have been obtained in this
paper from the consideration of extreme momentum configurations.
(A rigorous derivation of  (\ref{curve1}) and (\ref{curve2}) will
be provided in the forthcoming paper \cite{cho}.)
 Fitting the data points with the curves (\ref{curve1})
and (\ref{curve2}), we obtain $m_{\tilde g} = 776.3 \pm 1.3$ GeV
and $m_{\tilde\chi_1^0}=97.3 \pm 1.7$ GeV, which are quite close
to the true values, $m_{\tilde g}=780.3$ GeV and
$m_{\tilde\chi_1^0}=97.9$ GeV. This demonstrates that the gluino
stransverse mass can be very useful for measuring the gluino and
the LSP masses experimentally.

%
%

Let us now consider the case that squark mass $m_{\tilde q}$ is
smaller than the gluino mass $m_{\tilde g}$. In such case, the
following cascade decay is open;
\begin{eqnarray}
\tilde g \rightarrow q \tilde q \rightarrow q q {\tilde\chi_1^0}
\label{eq:2body}.
\end{eqnarray}
In this case also,  we consider two extreme momentum configurations
which are similar to those considered for three body gluino decay,
and construct the corresponding gluino stransverse masses. Here
again, we assume gluinos are pair produced at rest and each gluino
decays into a quark and a squark on the transverse plane with
respect to the proton beam direction.

If the quark from squark decay is produced in the same direction as
the first quark from gluino decay, and the two sets of gluino decay
products are parallel to each other, the transverse energies and
transverse momenta of the $qq$ systems are given by
\begin{eqnarray}
 &&E_T^{qq(1)}=E_T^{qq(2)}
= |\bold{p}_T^{qq(1)}| =|\bold{p}_T^{qq(2)}|\nonumber \\
&&= {m_{\tilde g}^2 -m_{\tilde\chi_1^0}^2 \over 2 m_{\tilde g}}\,
\equiv \, E_T^{qq}(max),\nonumber \label{etmax}
\end{eqnarray}
and we have $m_{qqT}^{(1)}=m_{qqT}^{(2)}$ for the transverse
invariant masses of the $qq$ systems and $|\bold{p}_T^{miss}|=2
E_T^{qq} (max)$ for the total observed missing transverse momentum.
Then the same procedure to obtain the gluino stransverse mass
(\ref{eq:mg1}) can be applied to this case, leading to $m_{T2}^{\rm
max}$ which is same as Eq.(\ref{curve1}):
\begin{eqnarray}
m^{\rm max}_{T2}(m_\chi)&=& {m_{\tilde g}^2 -m_{\tilde\chi_1^0}^2
\over 2 m_{\tilde g}} + \sqrt{\left({m_{\tilde g}^2
-m_{\tilde\chi_1^0}^2
\over 2 m_{\tilde g}}\right)^2 + m_\chi^2} \nonumber \\
&&\mbox{for}\quad m_\chi\leq m_{\tilde\chi_1^0} .
\label{curve3}\end{eqnarray}

Now we consider other extreme momentum configuration in which the
quark from squark decay is produced in the opposite direction to
the first quark from gluino decay. In this case, the invariant
transverse masses, transverse energies and transverse momenta of
two $qq$ systems are given by
\begin{eqnarray}
m_{qqT}^{2(1)}&=&m_{qqT}^{2(2)} \,=\, {(m_{\tilde g}^2 - m_{\tilde
q}^2) (m_{\tilde q}^2 - m_{\tilde\chi_1^0}^2)
\over m_{\tilde q}^2}, \nonumber\\
E_T^{qq(1)}&=&E_T^{qq(2)} \,=\, {m_{\tilde g} \over 2} (1-{m_{\tilde
q}^2 \over m_{\tilde g}^2})
+{m_{\tilde g} \over 2} (1-{m_{\tilde\chi_1^0}^2 \over m_{\tilde q}^2}),
\nonumber \\
|{\bold p}_T^{qq(1)} &=&{\bold p}_T^{qq(2)}|\,=\, |{m_{\tilde g}
\over 2} (1-{m_{\tilde q}^2 \over m_{\tilde g}^2}) -{m_{\tilde g}
\over 2} (1-{m_{\tilde\chi_1^0}^2 \over m_{\tilde q}^2})|, \nonumber
\end{eqnarray}
with the total missing transverse momentum given by ${\bold
p}_T^{miss} = -2\ {\bold p}_T^{qq(1)}$.

Imposing the two constraints, $m_T^{(1)}=m_T^{(2)}$ and
$\bold{p}_T^{miss}=\bold{p}_T^{\chi(1)}+\bold{p}_T^{\chi(2)}$, on
the momentum configurations having $m_{qqT}^{(1,2)}$, $E_T^{qq(1,2)}$, 
${\bold p}_T^{qq(1,2)}$, 
${\bold p}_T^{miss}(= -2\ {\bold p}_T^{qq(1)})$ as above, we obtain the
following balanced solution of $m_{T2} (\tilde g)$ at
$\bold{p}_T^{\chi(1)}=\bold{p}_T^{\chi(2)}=-\bold{p}_T^{qq(1)}$:
\begin{eqnarray}
m_{T2}^2 (\tilde g) &=& m_{qqT}^{2(1)} + m_\chi^2 \\
&+& 2 (E_T^{qq(1)} \sqrt{|{\bold p_T^{qq(1)}}|^2+m_\chi^2}+|{\bold
p_T^{qq(1)}|^2}). \nonumber \label{mt2-22}
\end{eqnarray}
Again,  one can show that this represents $m^{\rm max}_{T2}$ for
$m_\chi\geq m_{\tilde\chi_1^0}$ \cite{cho}, yielding
\begin{eqnarray}
m^{\rm max}_{T2}= \left({m_{\tilde g} \over 2} (1-{m_{\tilde q}^2
\over m_{\tilde g}^2}) +{m_{\tilde g} \over 2}
(1-{m_{\tilde\chi_1^0}^2 \over m_{\tilde q}^2})\right)\nonumber \\
+\sqrt{\left({m_{\tilde g} \over 2} (1-{m_{\tilde q}^2 \over
m_{\tilde g}^2}) -{m_{\tilde g} \over 2} (1-{m_{\tilde\chi_1^0}^2
\over m_{\tilde q}^2})\right)^2+m_\chi^2}.\label{curve4}
\end{eqnarray}
Note that the two functions (\ref{curve3}) and (\ref{curve4}) cross
at $m_\chi=m_{\tilde\chi_1^0}$ for which $m_{T2}^{\rm
max}=m_{\tilde{g}}$.

\begin{figure}[ht!]
\begin{center}
\epsfig{figure=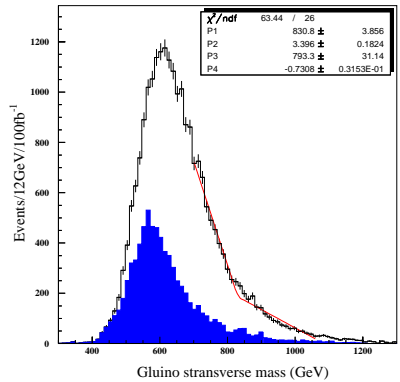,width=6cm,height=6cm}
\end{center}
\caption{\it The $m_{T2} (\tilde g)$ distribution with $m_\chi=350$
GeV for the benchmark point of mirage mediation. }
\label{fig:mirage1}
\end{figure}

\begin{figure}[ht!]
\begin{center}
\epsfig{figure=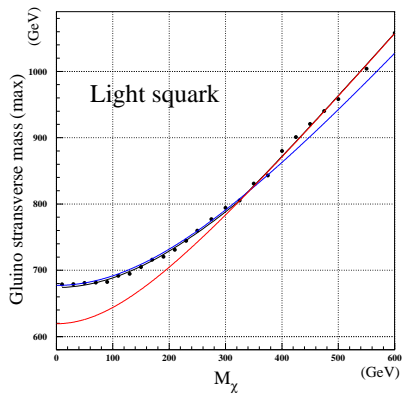,width=6cm,height=6cm}
\end{center}
\caption{\it $m_{T2}^{\rm max}$ as a function of the trial LSP mass
$m_\chi$ for the benchmark point of mirage mediation.}
\label{fig:mirage2}
\end{figure}

If one could construct (\ref{curve3}) and (\ref{curve4})
accurately enough from data, one would be able to determine all
involved sparticle masses, i.e. $m_{\tilde{g}},
m_{\tilde\chi_1^0}$ and $m_{\tilde{q}}$. To see how feasible it
is, we examined a parameter point of mirage mediation model
\cite{mirage}, providing the following sparticle masses;
\begin{eqnarray}
m_{\tilde g}=821.4\ {\rm GeV}, ~m_{\tilde q}=694.0\ {\rm GeV},~m_{\tilde\chi_1^0} = 344.2\ {\rm GeV}.
\nonumber
\end{eqnarray}
%
We have generated a Monte Carlo sample for this benchmark point of
mirage mediation, corresponding to $100 fb^{-1}$ integrated
luminosity. After the event selection cuts similar to the case of
three body gluino decay, we obtain  Fig.\ref{fig:mirage1} showing
the distribution of $m_{T2}(\tilde{g})$ for the trial LSP mass
$m_\chi=350$ GeV.
The edge value $m_{T2}^{\rm max}$ as a function of $m_\chi$ is
shown in Fig.\ref{fig:mirage2}. Fitting the data points to  the
curves (\ref{curve3}) and (\ref{curve4}),  we obtain $m_{\tilde
g}=799.5\pm 11.1$ GeV, $m_{\tilde q}=678.2\pm 7.0$ GeV and
$m_{\tilde\chi_1^0}=316.7\pm 15.4$ GeV. Though the overall scale
of the fitted values are well close to the true sparticle masses,
the central values are somewhat lower than the true values, which
is mainly due to a mild crossing of two curves.
The situation can be improved if we include the information from
{\it squark stransverse mass} for the process $pp \rightarrow
\tilde q \tilde q \rightarrow q\tilde\chi_1^0 q\tilde\chi_1^0$,
providing a relation between the edge value of the squark
stransverse mass and the trial LSP mass \cite{cho}:
\begin{eqnarray}
m_{T2}^{\rm max}(\mbox{squark})={m_{\tilde q}^2-m_{\tilde\chi_1^0}^2
\over 2m_{\tilde q}}+\sqrt{\left({m_{\tilde
q}^2-m_{\tilde\chi_1^0}^2 \over 2m_{\tilde
q}}\right)^2+m_\chi^2}.\nonumber
\end{eqnarray}
Including such information, we get $m_{\tilde g}=803.4 \pm 6.0$ GeV
and $m_{\tilde\chi_1^0}=322.4\pm 7.7$ GeV for the benchmark point.
The discrepancy between the fitted mass values and the true mass
values may still come from various systematic uncertainties such as
the effects of event selection cuts,  which  is beyond the scope of
this paper.


%
%

To conclude, we have introduced the gluino stransverse mass, and
shown that it can be used to determine the gluino mass and the
lightest neutralino mass separately, and also the 1st and 2nd
generation squark masses if squarks are lighter than gluino.

\vspace{0.3cm} This work was supported by the BK21 program of
Ministry of Education (WSC,KC,CBP), the Astrophysical Research
Center for the Structure and Evolution of the Cosmos funded by the
KOSEF (YGK), the KRF Grant KRF-2005-210-C000006 funded by the
Korean Government and the Grant No. R01-2005-000-10404-0 from the
Basic Research Program of the Korea Science $\&$ Engineering
Foundation.

\vfill\eject
\newpage

\begin{thebibliography}{99}

\bibitem{atlas}
ATLAS Technical Proposal, CERN-LHCC-94-43.

\bibitem{cms}
CMS Physics Technical Design Report, CERN-LHCC-2006-021.

\bibitem{susy}
H. P. Nilles, Phys. Rept. {\bf 110} (1984) 1; H. E. Haber and G. L.
Kane, Phys. Rept. {\bf 117} (1985) 75.

\bibitem{choi-nilles}
K. Choi and H. P. Nilles, JHEP {\bf 0704} (2007) 006.

\bibitem{lester}
C. G. Lester and D. J. Summers, Phys. Lett. {\bf B 463} (1999) 99;
A. Barr, C. Lester, and P. Stephens, J. Phys. {\bf G 29} (2003)
2343; C. Lester and A. Barr, arXiv:0708.1028.


\bibitem{cho} W. S. Cho, K. Choi, Y. G. Kim and C. B. Park, in
preparation.


\bibitem{amsb} L. Randall and R. Sundrum, Nucl. Phys. {\bf B 557}
(1999) 79 [hep-th/9810155]; G. F. Giudice, M. A. Luty, H. Murayama
and R. Rattazzi, JHEP {\bf 12} (1998) 027 [hep-ph/9810442].

\bibitem{pythia}
 T. Sjostrand, P. Eden, C. Friberg, L. Lonnblad, G. Miu, S. Mrenna and
 E. Norrbin, Computer Physics Commun. 135 (2001) 238;
 T. Sjostrand, S. Mrenna and P. Skands,
 LU TP 06-13, FERMILAB-PUB-06-052-CD-T [hep-ph/0603175].




\bibitem{pgs}
http://www.physics.ucdavis.edu/\~ conway/research/
software/pgs/pgs4-general.htm.

\bibitem{mirage}
K. Choi, A. Falkowski, H. P. Nilles and M. Olechowski, Nucl. Phys.
{\bf B718} (2005) 113 [hep-th/0503216];
  K.~Choi, K.~S.~Jeong and K.~i.~Okumura,
  JHEP {\bf 0509}, 039 (2005)
  [arXiv:hep-ph/0504037].


\end{thebibliography}
\end{document}